\documentclass[usenatbib]{mn2e}
\usepackage{multirow}
\usepackage{rotating}
\usepackage{colortbl}
\usepackage{color}
\usepackage[fleqn]{amsmath}
\usepackage{amssymb}
\usepackage{amsfonts}
\usepackage{verbatim}
\usepackage{scalefnt}
\usepackage[percent]{overpic}

\def\gtorder{\mathrel{\raise.3ex\hbox{$>$}\mkern-14mu
     \lower0.6ex\hbox{$\sim$}}}
\def\ltorder{\mathrel{\raise.3ex\hbox{$<$}\mkern-14mu
     \lower0.6ex\hbox{$\sim$}}}

\newcommand{\msun}{\,{\rm M_\odot}}

\newcommand{\edd}{{\rm Edd}}

\newcommand{\beq}{\begin{equation}}
\newcommand{\eeq}{\end{equation}}
\newcommand{\ba}{\begin{eqnarray}}
\newcommand{\ea}{\end{eqnarray}}
\def\spose#1{\hbox to 0pt{#1\hss}}
\newcommand{\lta}{\mathrel{\spose{\lower 3pt\hbox{$\mathchar"218$}}
      \raise 2.0pt\hbox{$\mathchar"13C$}}}
\newcommand{\gta}{\mathrel{\spose{\lower 3pt\hbox{$\mathchar"218$}}
      \raise 2.0pt\hbox{$\mathchar"13E$}}}

\newcommand{\comments}[1]{} 

\title[Magnetically supported disks in AGN]{Magnetically elevated accretion disks in active galactic nuclei: broad emission line regions and associated star formation}

\author[Begelman  and Silk]{Mitchell C. Begelman,$^{1,2}$ Joseph Silk$^{3,4,5,6}$\\
$^{1}$ JILA, University of Colorado and NIST, 440 UCB, Boulder, CO 80309-0440, USA\\
$^{2}$ Department of Astrophysical and Planetary Sciences, 391 UCB, University of Colorado, Boulder, CO 80309-0391, USA\\
$^{3}$ Institut d'Astrophysique de Paris, Sorbonne Universit\'es, UPMC Univ. Paris 06 et CNRS, UMR 7095, F-75014, Paris, France\\
 $^{4}$ Laboratoire AIM-Paris-Saclay, CEA/DSM/IRFU, CNRS, Universit\'e Paris Diderot,  F-91191 Gif-sur-Yvette, France\\
$^{5}$ Department of  Physics \& Astronomy, The Johns Hopkins University, Baltimore, MD 21218, USA\\
$^{6}$ Beecroft Institute of Particle Astrophysics and Cosmology, Department of Physics,
University of Oxford,  Oxford OX1 3RH, UK
}

\begin{document}

\maketitle

\begin{abstract}
We propose that the accretion disks fueling active galactic nuclei are supported vertically against gravity by a strong toroidal ($\phi-$direction) magnetic field that develops naturally as the result of an accretion disk dynamo.  The magnetic pressure elevates most of the gas carrying the accretion flow at $R$ to large heights $z \gta 0.1 R$ and low densities, while leaving a thin dense layer containing most of the mass --- but contributing very little accretion --- around the equator.  We show that such a disk model leads naturally to the formation of a broad emission line region through thermal instability.  Extrapolating to larger radii, we demonstrate that local gravitational instability and associated star formation are strongly suppressed compared to standard disk models for AGN, although star formation in the equatorial zone is predicted for sufficiently high mass supply rates.  This new class of accretion disk models thus appears capable of resolving two longstanding puzzles in the theory of AGN fueling: the formation of broad emission line regions and the suppression of fragmentation thought to inhibit accretion at the required rates. We show that the disk of stars that formed in the Galactic Center a few million years ago could have resulted from an episode of magnetically elevated accretion at $\gta 0.1$ of the Eddington limit.

\end{abstract}


\section{Introduction}\label{sec:Introduction}

Simulations of accretion disks, both local \citep[in the ``shearing box" approximation: e.g.,][]{brandenburg95,davis10,simon12} and global \citep{beckwith11,oneill11}, have long suggested that the operation of the magnetorotational instability (MRI) --- essential for local angular momentum transport \citep{balbus98} --- can also lead to the generation of a large-scale toroidal magnetic field through a dynamo process.  The strength of the dynamo field is enhanced by the presence of net magnetic flux (vertical field) threading the disk, which also enhances the rate of angular momentum transport \citep{hawley95}. In any case, the resulting toroidal field is much stronger than the imposed or stochastic poloidal field.  Recent shearing box simulations \citep{bai13,salvesen16a} have shown that a vertical field with a magnetic pressure exceeding about $0.1\%$ of the central gas pressure leads to a dynamo field whose pressure dominates the disk everywhere.  We call such disks, which are supported against the vertical component of gravity by magnetic pressure, {\it magnetically elevated disks}.\footnote{Magnetically elevated disks are distinct from ``magnetically arrested disks" \citep[MAD:][]{narayan03,igumenshchev08}, which require much larger poloidal fluxes.}

The implications of magnetic support for accretion disk phenomenology are profound.  Magnetically elevated disks are expected to be much thicker and, for the most part, less dense than standard accretion disks.  Consequently, they should be less susceptible to gravitational instability and fragmentation into stars \citep{pariev03,begelman07,gaburov12}, resolving a major stumbling block to the application of standard accretion disk theory to active galactic nuclei (AGN) \citep{kolykhalov80,shlosman87,shlosman89a,goodman03}.  Their larger scale heights, faster inflow rates, and larger color corrections are in line with a variety of observations of X-ray binaries and cataclysmic variables \citep{begelman07}, while the coupling between poloidal flux and disk structure suggests intriguing new possibilities for explaining state transitions and magnetized winds \citep{begelman14,begelman15}.

It is somewhat surprising that disks should be able to maintain the strong levels of magnetization seen in simulations.  Horizontal magnetic fields are buoyant and should be continuously escaping, as simulations indeed show \citep{miller00}. The structure and energetics of magnetically elevated disks can be modeled analytically in terms of a competition between the creation of toroidal field by MRI and its loss through buoyancy and work done on the gas while escaping \citep{begelman15}.   Because the field is continuously replenished by the dynamo process associated with MRI, which operates far from the equatorial plane, the magnetic pressure $p_{\rm B}$ in an elevated disk declines very gradually with height, while the density drops off rapidly outside a narrow core region where it is approximately uniform.   This means that a disk that is moderately magnetically dominated near the midplane, becomes overwhelmingly dominated by magnetic pressure at a few scale heights above the core region.    Although most of the mass in such a disk is confined to the equatorial layer, most of the dissipation and resulting accretion occur far from the midplane, where the density is very low.  Shearing box simulations appear to agree qualitatively, and to some extent quantitatively, with this model \citep{salvesen16a}.

One aspect of the model that has not been tested numerically is the extent to which MRI continues to operate far from the equator.  Applications of the model to date \citep{begelman07,begelman15} have adopted the criterion derived by \cite{pessah05}, which states that MRI switches off when the Alfv\'en speed $v_{\rm A\phi}$ associated with the toroidal field exceeds the geometric mean of the Keplerian speed and the gas sound speed.  This limits the disk height to $H \sim (c_{\rm s}/ v_{\rm K})^{1/2} R$, where $c_{\rm s}$ is the gas sound speed and $v_{\rm K}$ is the Keplerian speed, respectively.  This limit, however, was derived from a local linear stability analysis assuming no radial or vertical stratification of the magnetic pressure or any other fluid quantity.  These must be present if the disk is magnetically supported.  In the absence of results incorporating global and nonlinear effects, we will parametrize our ignorance by treating $H/R \equiv \xi$ as a constant.  

In this paper we apply the \cite{begelman15} model for the structure of magnetically elevated disks to the fueling of AGN. In section 2 we review the model and our current best estimates of parameters needed for the application, based on existing local simulations.  In section 3 we show that such disks can undergo thermal instability, conducive to the formation of broad emission line regions with the observed properties.  
We discuss disk self-gravity in section 4, where we estimate threshold mass supply rates for fragmentation to set in, and star formation rates if it does.  We argue that the unique properties of magnetically elevated disks may lead to substantial star formation coexisting with accretion in luminous AGN.  We discuss our results and conclusions in section 5.    
   
\section{Magnetically elevated disk models}\label{sec:sec2} 

Conditions on the equatorial plane of a magnetically elevated disk consist of the central magnetic pressure, $p_{{\rm B}0}$, gas+radiation pressure, $p_0$, and density, $\rho_0$.  There is also the pressure of the poloidal magnetic field, $p_{\rm p} = B_{\rm z}^2/8\pi$, which is assumed to be independent of $z$.  The suite of simulations performed by \cite{salvesen16a} suggests that these parameters are connected by the approximate relation
\beq
\label{pbzero}
p_{{\rm B}0} \sim 10 (p_0 p_{\rm p})^{1/2}
\eeq
for $p_{\rm p} \gta 10^{-5} p_0$.  Equivalently, the plasma beta parameter on the equator is given by
\beq
\label{betazero}
\beta_0 \equiv {p_0\over p_{{\rm B}0}} \sim 10^{-2} {p_{{\rm B}0}\over p_{{\rm p}}}.
\eeq
Even if magnetic pressure strongly dominates on the equator ($\beta_0 \ll 1$), the density and gas pressure scale height on the equatorial plane, $H_0$, is given by its usual value in the absence of magnetic support.   Anticipating that gas pressure dominates over radiation pressure on the central plane of even the most luminous magnetically elevated disks, we have 
\beq
\label{Hzero}
H_0 \sim \theta_0^{1/2} x^{3/2} r_{\rm g} 
\eeq
where $\theta_0 \equiv kT_0/\mu c^2$ for a central temperature $T_0$ and mean particle mass $\mu$ (taken to be $0.6 m_{\rm p}$ in our applications), $r_{\rm g} \equiv GM/c^2 = 1.5 \times 10^{13} m_8$ cm is the gravitational radius of an $M = 10^8 m_8 \msun$ black hole, and $x\equiv R/r_{\rm g}$. 

According to the \cite{begelman15} model, the magnetic pressure varies with height according to 
\beq
\label{pbzero3}
p_{\rm B} \approx  p_{{\rm B}0} \left( 1 + {z^2\over 4 H_0^2} \right)^{-\beta_0/(1+ \beta_0)} \approx  p_{{\rm B}0} \left({z\over 2 H_0} \right)^{-2\beta_0/(1+ \beta_0)}
\eeq
for $z \gg H_0$. In addition to any constraint due to the strong toroidal field --- which we have parametrized by assuming that MRI persists up to $z \sim \xi R$ ---  MRI is also subject to quenching at the original \cite{balbus91} limit, $k_{\rm z} v_{\rm Az} > \sqrt{3}$, where $k_{\rm z}$ is the vertical wavenumber and $v_{\rm Az} = (2 p_{\rm p}/\rho)^{1/2}$ is the Alfv\'en speed associated with the poloidal field.  At $z \gg H_0$, this condition is effectively equivalent to $p_{\rm p} \gta p_{\rm B}$, which is never satisfied if $\beta_0 \ll 1$.  On the other hand, MRI will be quenched in the equatorial layer, $z \lta H_0$, if $p_{\rm p} > p_0$, corresponding to $\beta_0 < 0.1$ (see equation \ref{betazero}). Because the disk is geometrically thick and MRI-active throughout, it is reasonable to suppose that it is relatively effective in dragging in magnetic flux \citep{lubow94,guilet12}, and we will therefore assume that $\beta_0$ tends to evolve toward this minimum value.  That is to say, in our quantitative estimates below we will assume that $\beta_0 \approx 0.1$ at all radii, although we note that most numerical values are insensitive to the value of $\beta_0$ provided it is $\ll 1$.

We set the flux dissipated on one side of the equator to the energy liberated by accretion:
\beq
\label{Fd}
F_{\rm d} \approx  \alpha \Omega \int^{\xi R}_0 p_{\rm B} dz \approx {3\over 8\pi}  {GM\dot M \over R^3},
\eeq
where $\alpha$ is the \cite{shakura73} viscosity parameter, plausibly $\sim 0.3$ in this highly magnetized limit \citep{salvesen16a} and $\Omega$ is the Keplerian angular velocity. Setting $\beta_0 = 0.1$, we solve equation (\ref{Fd}) to obtain
\beq
\label{pbzero2}
p_{{\rm B}0} \approx  3.0\times 10^9 \xi^{-0.8}\alpha_{0.3}^{-1} m_8^{-2}\dot m\theta_0^{-0.1}  x^{-2.6} \ {\rm erg \ cm}^{-3},
\eeq
where $\dot m = \dot M / 1  M_\odot$ yr$^{-1}$. The central density is given by 
\beq
\label{rhozero2}
\rho_0 = {\beta_0 p_{{\rm B}0}\mu \over k T_0} \approx  3.3\times 10^{-13} \xi^{-0.8}\alpha_{0.3}^{-1} m_8^{-2}\dot m\theta_0^{-1.1}  x^{-2.6} \ {\rm g \ cm}^{-3}.
\eeq

While the equatorial conditions depend on the thermodynamics of the flow through $\theta_0$, conditions at $z \sim H$, where most of the accretion takes place, are independent of this parameter.  Using equation (\ref{pbzero3}), we obtain
\beq
\label{pbH}
p_{\rm B} (H) \approx  3.4\times 10^9 \xi^{-1}\alpha_{0.3}^{-1} m_8^{-2}\dot m x^{-2.5} \ {\rm erg \ cm}^{-3}
\eeq
and
\begin{eqnarray}
\label{rhoH}
\rho (H) & =  & - {1 \over \Omega^2 H} {dp_{\rm B}\over dz}(H) \nonumber\\ 
& \approx & 6.9\times 10^{-13} \xi^{-3}\alpha_{0.3}^{-1} m_8^{-2}\dot m  x^{-1.5} \ {\rm g \ cm}^{-3}.
\end{eqnarray}

\section{The broad emission line region}\label{sec:sec3}

Broad emission lines (BELs) in AGN are thought to arise from clouds or filaments of photoionized gas with a small volume filling factor but large covering fraction $\gta 0.1$ and substantial column density ($\sim 10^{23}$ cm$^{-2}$), located at distances $x \sim 10^2 - 10^5$ from the black hole \citep{davidson72}.  Theoretical photoionization studies have shown that there is a strong observational selection effect for BELs to form under ``optimally emitting" conditions of particle density and ionizing photon flux \citep{baldwin95}, which translate into a tight correlation between distance from the black hole and luminosity.  The theoretical question of why AGN have broad-line regions (BLRs) thus boils down to whether filaments with the optimal density, covering factor and column density exist at the right distance.  In this section we argue that magnetically elevated disks will naturally produce such filaments through a thermal instability.    

Optimally emitting BEL clouds are characterized by densities $n \sim 10^{10} n_{10}$ cm$^{-3}$ and a broader range of dimensionless ionization parameters
\beq
\label{Uion}
U \equiv {\dot N_{\rm ion} \over 4\pi R^2 c n } 
\eeq
with typical values $U \sim 10^{-2}$, where $\dot N_{\rm ion}$ is the production rate of photons capable of ionizing hydrogen.  It is difficult to measure $\dot N_{\rm ion}$ directly, but reverberation studies of the broad H$\beta$ line give a dimensionless radius of the BLR,
\beq
\label{xBLR}
x_{\rm BLR} \approx  6.2\times 10^3 L_{45}^{1/2} m_8^{-1} 
\eeq
\citep{bentz09}, where $10^{45}L_{45}$  erg s$^{-1}$ is the bolometric luminosity and we adopt the bolometric correction $k_{{5100}\AA} = 8.1$ from \cite{runnoe12}.   

Low-density gas exposed to the AGN continuum will equilibrate to the inverse Compton temperature 
\beq
\label{TIC}
T_{\rm IC} = {1\over 4 k u} \int u_\nu h\nu d\nu = 10^7 T_{\rm IC7} ,
\eeq
where $u$ and $u_\nu$ are the total and spectral radiation energy density, respectively.  But if the density becomes high enough, two-body cooling processes will drive the gas thermally unstable, so that a fraction of it will cool to $\sim 10^4$ K and form a two-phase medium capable of producing strong UV and optical emission lines \citep{mccray79,krolik81}. 

For $T_{\rm IC7} \gta 1$ the principal collisional cooling mechanism is thermal bremsstrahlung, and the condition for isobaric thermal instability can be written  
\beq
\label{xth}
{\rho \over u} >  3.2 \times 10^{-17} T_{\rm IC7}^{1/2} 
\eeq
\citep{begelman90}, where $\rho$ and $u$ are in cgs units.  Using $\rho(H)$ from equation (\ref{rhoH}) and assuming an accretion efficiency $0.1 \epsilon_{-1}$, this implies that the uppermost layers of the magnetically elevated disk should go thermally unstable for 
\beq
\label{xth2}
x > x_{\rm th} \approx  1.0\times 10^{7} \xi^6 (\epsilon_{-1} \alpha_{0.3})^2 T_{\rm IC7} . 
\eeq
A necessary condition for the formation of a BLR is then that $x_{\rm BLR} > x_{\rm th}$ or, equivalently, 
\beq
\label{xicond}
\xi < 0.3 (\epsilon_{-1} \alpha_{0.3})^{-1/3} (m_8 T_{\rm IC7})^{-1/6}  L_{45}^{1/12}. 
\eeq
This condition is extremely insensitive to all parameters, and is readily satisfied for elevated disk models truncated at $\xi \gta 0.1$, which also provide adequate covering fraction to produce the observed line intensities.  Given that at least half the mass in the two-phase regions should be in line-emitting clouds, we obtain column densities    
\beq
\label{NBLR}
N_{\rm BLR} > 9.3 \times 10^{23}   T_{\rm IC7}^{1/2}  L_{45}^{1/2} \ {\rm cm}^{-2}, 
\eeq
which are also adequate to produce the observed lines.

Finally, we check the ability of the hot phase to confine the line-emitting gas at densities inferred to exist in the BLR.  Assuming that the two-phase system remains close to the thermal stability threshold (\ref{xth}), we obtain a hot-phase pressure given by
\beq
\label{phot}
p_{\rm hot} \approx 1.5 \times 10^{-18}   T_{\rm IC7}^2  u_{\rm BLR}^2 \rho_{\rm hot}^{-1} \ {\rm erg \ cm}^{-3}, 
\eeq
where both the density of the hot phase, $\rho_{\rm hot}$, and the radiation energy density at $x_{\rm BLR}$, $u_{\rm BLR} \approx 0.3$ erg cm$^{-3}$, are expressed in cgs units. Since $\rho_{\rm hot} < \rho (H)$ measured at $x_{\rm BLR}$, this gives a lower limit on the hot phase pressure, 
\beq
\label{phot2}
p_{\rm hot} > 0.6  \xi^3 \epsilon_{-1} \alpha_{0.3} m_8^{1/2} T_{\rm IC7}^2 L_{45}^{-1/4}  \ {\rm erg \ cm}^{-3} . 
\eeq
If the line-emitting gas, with a temperature $10^4 T_4$ K, is in pressure balance with the hot phase this implies a lower limit on the particle density,
 \beq
\label{phot3}
n  > 4.5 \times 10^{11}\xi^3 \epsilon_{-1} \alpha_{0.3} m_8^{1/2} T_{\rm IC7}^2 T_4^{-1} L_{45}^{-1/4}  \ {\rm cm}^{-3}, 
\eeq  
consistent with the inferred densities of the broad-line gas if $\xi \gta 0.1$.

\section{Self-gravity and star formation}\label{sec:sec4} 

We next extrapolate the elevated-disk model to radii where self-gravity and fragmentation of the disk becomes an issue in standard accretion disk models of AGN.  Low gas density due to magnetic support is the key idea behind earlier suggestions that magnetically supported disks could evade self-gravity in AGN \citep{pariev03,begelman07,gaburov12}.  What is new is the realization that the density in a magnetically elevated disk should be strongly stratified with height, with accretion at high $z$ coexisting with a dense, equatorial layer where fragmentation and star formation may well occur. Our objective in this section is to assess the connections between these regions.   

Strong density stratification means that fragmentation and star formation, when it occurs at all, should be dominated by the equatorial layer.  Therefore, we estimate the Toomre Q-parameter using the conditions on the midplane.  The condition for gravitational instability is 
\beq
\label{Qzero}
Q_0 \sim {M \over 2 \pi \rho_0 R^3}  < Q_{\rm  crit} ,  
\eeq
where $Q_{\rm crit} =1$ for a fluid dynamical disk but may differ in the presence of a strong toroidal field. (Riols \& Latter [2016] find that $Q_{\rm crit}$ may be as large as $\sim 10$ in two-dimensional MHD simulations of gravitoturbulent shear flows in which MRI is suppressed.) The value of $Q_0$ depends on the temperature of the equatorial layer, which could be affected by external irradiation \citep{shlosman87} and other heating mechanisms \citep[][and references therein]{ginsburg16}.  Following \cite{shlosman87} we assume that the equatorial layer comes into equilibrium with a fraction $\chi = 10^{-2} \chi_{-2}$ of the AGN flux $L/4\pi r^2$.  Since the cooling timescale is shorter than $3/\Omega$ \citep{gammie01}, instability leads to fragmentation at radii  
\begin{eqnarray}
\label{xfrag}
x & > & x_{\rm frag}  \nonumber \\
& \approx &   2.1 \times 10^{5} \xi^{16/19} \left({\alpha_{0.3} \over Q_{\rm crit} }\right)^{20/19} m_8^{-11/19} \dot m ^{-29/38} , 
\end{eqnarray}
with a threshold temperature 
\beq
\label{Tfrag}
T(x_{\rm frag}) \approx  310 \xi^{-8/19} \left({\alpha_{0.3} \over Q_{\rm crit} }\right)^{-10/19} m_8^{-4/19}\dot m^{12/19} \ {\rm K} ,
\eeq
where we have suppressed weak dependences $\propto (\chi_{-2}\epsilon_{-1})^{11/38}$ and $\propto (\chi_{-2}\epsilon_{-1})^{2/19}$ in equations (\ref{xfrag}) and (\ref{Tfrag}), respectively.  
 
This estimate is valid inside the black hole's gravitational radius of influence $r_{\rm BH}$ within its host galaxy.  Modeling the stellar potential of the galactic nucleus as that of an isothermal sphere with velocity dispersion $\sigma = 200 \sigma_{200}$ km s$^{-1}$, we have 
\begin{equation}
\label{RBH}
x_{\rm BH} = \left({c \over  \sigma }\right)^2 = 2.3 \times 10^6 \sigma_{200}^{-2} .
\end{equation}
Outside $r_{\rm BH}$, the enclosed mass increases linearly with $R$, and the equatorial density of a magnetically elevated disk carrying a fixed mass flux scales as $\rho_0 \propto R^{-2} T^{-1.1}$.  (We note that $x = R/r_{\rm g}$ is effectively constant outside $r_{\rm BH}$, since $r_{\rm g}$ is proportional to the enclosed mass.)  This means that $Q_0 \propto T^{1.1}$ with no explicit radial dependence, so the susceptibility to fragmentation increases outside $r_{\rm BH}$ only if the temperature continues to drop.  However, if the temperature levels off to a constant value $T = 100 T_2$ K at $R \lta r_{\rm BH}$, then no fragmentation will occur, at any radius, if $R_{\rm frag} > r_{\rm BH}$, corresponding to a lower limit on the local accretion rate needed to trigger fragmentation,
\beq
\label{mdotmax}
\dot M > 0.1 \xi^{4/5} \left({\alpha_{0.3} \over Q_{\rm crit} }\right) \sigma_{200}^{4/5} T_2^{11/10}\ \msun \ {\rm yr}^{-1},
\eeq
and an associated AGN luminosity
\beq
\label{Lmax}
L_{45} >  0.62 \xi^{4/5} \left({\epsilon_{-1}\alpha_{0.3} \over Q_{\rm crit} }\right) \sigma_{200}^{4/5}T_2^{11/10}.
\eeq
We stress that this is the threshold for mass flux carried by the entire thick disk, not just the flux carried by the thin equatorial layer.   We also note that this condition applies to the mass flux passing through the region around $r_{\rm BH}$, which may be higher than the accretion rate reaching the black hole (in which case the luminosity in equation \ref{Lmax} would be lower).  
   
If fragmentation starts in the equatorial layer, we assume that it proceeds all the way to star formation at a rate 
\beq
\label{Sigmadot}
\dot \Sigma_* = \epsilon_* \Omega \Sigma_0 
\eeq
per unit area, where $\Sigma_0 = 2\rho_0 H_0$ is the surface density of the equatorial layer and $\epsilon_*$ is the star formation efficiency.  It is reasonable to suppose that feedback from massive stars drives turbulence which regulates the density at close to the critical value for fragmentation, implying $\rho_0 \sim \Omega^2 / (2\pi G Q_{\rm crit})$ and $H_0 \sim v_{\rm t}/\Omega$, where $v_{\rm t}$ is the turbulent velocity dispersion.  We assume that the MRI-driven dynamo continues to operate in the presence of this turbulence, but with the toroidal magnetic field now governed by the turbulent pressure, 
\beq
\label{pBturb}
p_{{\rm B}0} \sim  \beta_0^{-1}\rho_0 v_{\rm t}^2 \sim {M v_{\rm t}^2\over 2\pi \beta_0 Q_{\rm crit} R^3} .
\eeq  
We can then use equation (\ref{pbzero2}) to express $v_{\rm t}$ in terms of the accretion rate $\dot M$, which is mostly flowing outside the turbulent star-forming layer:
\beq
\label{vturb}
v_{\rm t} \sim 3.1 (\xi \sigma_{200})^{-0.36} \left(\dot m {Q_{\rm crit}\over \alpha_{0.3}}{\beta\over 0.1} \right)^{0.45} \left( {R \over r_{\rm BH}}\right)^{0.18} \ {\rm km \ s}^{-1}.
\eeq   

We estimate the star formation efficiency by equating the energy injection rate by supernova remnants to the turbulent dissipation rate $\sim \Omega \Sigma_0 v_{\rm t}^2$.  The energy injection rate per unit area can be parametrized as 
\beq
\label{Edotturb}
\dot {\cal E}_{\rm SN} \sim \eta_{\rm SN} \dot \Sigma_* E_{\rm SN} f_{\rm SN} ,
\eeq 
where $\eta_{\rm SN} = 0.01 \eta_{-2} \msun^{-1}$ is the number of Type II supernovae per solar mass of star formation, $E_{\rm SN} = 10^{51}E_{51}$ erg is the initial kinetic energy injected by each supernova, and $f_{\rm SN}$ is the fraction of that kinetic energy going into turbulence.  In the limit of a weak magnetic field, energy injection occurs in the momentum-conserving limit when the blast waves have cooled \citep{hopkins11}.  For ambient densities in the expected range $\sim 10^{8}$ cm$^{-3}$, the expansion speed of the blast wave at the start of the snowplow phase (which turns out to be remarkably insensitive to ambient density) is $v_{\rm cool} \gta 10^3$ km s$^{-1}$, and $f_{\rm SN} \sim (v_{\rm t} / v_{\rm cool}) \lta 10^{-2}$ is the fraction of energy remaining when the typical speed decays to $v_{\rm t}$  \citep{kim15}.  However, in the presence of a strong ambient magnetic field (with $\beta_0 < 1$), most of the supernova expansion energy should go first into magnetic energy, which is harder to radiate away.  Thus, we expect $f_{\rm SN}$ to be much larger than the momentum-conserving limit, perhaps by an order of magnitude.  Writing $f_{\rm SN} = 0.1 f_{-1}$, we obtain a star formation efficiency 
\beq
\label{epsstar}
\epsilon_* \sim {v_{\rm t}^2  \over \eta_{\rm SN} E_{\rm SN} f_{\rm SN}} = 2 \times 10^{-3} {v_{\rm t,10}^2 \over \eta_{-2} E_{51} f_{-1}},
\eeq 
where $v_{\rm t,10} = v_{\rm t}/ 10$ km s$^{-1}$.  

Given the expressions for $v_{\rm t}$ and $\epsilon_*$, we can integrate the star formation rate per unit area to obtain the total star formation rate outside $R$, 
\beq
\label{Mdotstar}
\dot M_* (> R) = 2\pi \int_{R} \dot \Sigma_* R dR \sim {4.3 \epsilon_*(R)v_{\rm t}(R)M \over  Q_{\rm crit} R} \propto R^{-0.46} ,
\eeq 
where we have assumed a power-law behavior for $v_{\rm t} (R)$ according to equation (\ref{vturb}). We estimate the total star formation rate by evaluating equation (\ref{Mdotstar}) at $R_{\rm frag}$, with $v_{\rm t,10} = 0.012 T(x_{\rm frag})^{1/2}$ from equation (\ref{Tfrag}). The result is
\beq
\label{Mdotratio}
{\dot M_* \over \dot M} \sim 0.08 (\eta_{-2} E_{51} f_{-1})^{-1} \xi^{-28/19} \alpha_{0.3}^{-35/19} Q_{\rm crit}^{-5/19} m_8^{5/19} \dot m^{27/38} .  
\eeq 
Setting $\dot M_*/\dot M \sim 1$ effectively places an upper limit to the accretion rate able to reach the black hole --- all mass supplied in excess of this limit would go into star formation.  Given the fiducial values of the parameters, this limit is much higher than the limiting values predicted by standard accretion disk theory, and could plausibly support accretion in the most powerful known quasars with $\dot m$ of several hundred.  However, many of the parameters in this model are very uncertain, and it is important to recognize the effect on the predicted accretion and star formation rates if their values have been over- or underestimated.  In particular, larger values of $f_{\rm SN}$, $\alpha$ and $Q_{\rm crit}$ tend to suppress star formation, as does a top-heavy initial-mass function (IMF) leading to a larger value of $\eta_{\rm SN}$.  Such an IMF, with $\eta_{-2}$ as large as 10, has been inferred for the young stars in the Galactic Center, from X-ray observations \citep{nayakshin05}.  On the other hand, smaller values of the overall elevated disk thickness, as measured by $\xi = H/R$, would promote star formation and provide more stringent constraints on the mass flux reaching the black hole.


\section{Discussion and conclusions}\label{sec:sec5}

We have applied a simple model for the radial and vertical structure of magnetically elevated accretion disks \citep{begelman15} to conditions in AGN and have shown that it provides a natural explanation for the existence and properties of the broad emission line region, while simultaneously providing a framework for understanding the relationship between accretion and star formation in the outer disk. 

We argue that the BLR occurs as the result of thermal instability in the upper, accreting layers of the elevated disk, which are irradiated by the central engine.  The picture of a two-phase BLR established through thermal instability has been around for decades \citep{mccray79,krolik81}; what is new is the role of the magnetic field in levitating the gas to heights where it can intercept a sufficient fraction of the incident radiation, and in regulating its mean density.

It is important to note that the magnetic field is not primarily responsible for confining the line-emitting gas in this picture, in contrast to the model proposed by \cite{rees87} in which the BEL gas is fully confined by the pressure of the magnetic field.  Here, the line-emitting and Compton-heated phases are in pressure balance with each other, with regions of hot and cold gas strung out along the predominantly toroidal magnetic field lines.  Since the magnetic pressure is much larger than the thermal pressure, significant pressure fluctuations are possible transverse to the field, and corresponding density fluctuations have been seen in isothermal simulations of magnetically supported shearing boxes \citep{salvesen16a}.    

Slightly further from the black hole, where the effective temperature drops below the sublimation temperature ($\sim 1500$ K) for graphite, we expect the levitated gas to be dusty, with a covering factor and column densities consistent with the putative ``dusty torus" that reprocesses a significant fraction of the AGN continuum into the infrared.  However, in contrast to the BLR, gas in this region may be strongly susceptible to radiation pressure forces due to the incident UV flux, and any model of this region will have to take such forces into account \citep{dorodnitsyn12,chan16}. 

The same magnetically elevated disk models, extrapolated to still larger distances, provide a nuanced picture of the relationship between star formation and accretion in the AGN fueling process.  Standard thin disk models predict the onset of fragmentation, and interruption of the accretion flow, when $\dot M$ is larger than about $10^{-3} M_\odot$ yr$^{-1}$ \citep{shlosman87,goodman03}, corresponding to AGN luminosities $\lta 10^{43}$ erg s$^{-1}$. Simple one-zone models for magnetically supported disks, approximating $\rho \sim \rho(H)$ (equation \ref{rhoH}), yield drastically higher thresholds for fragmentation \citep[e.g.,][]{begelman07}, but fail to take into account the strong vertical density stratification associated with the buoyant escape of magnetic field \citep{begelman15}.  Incorporating this stratification, we find that the threshold for avoiding fragmentation altogether is not as high as in the one-zone case, but also that there is a new regime in which star formation in the equatorial layer coexists with unimpeded accretion at larger heights.  Only when the ratio $\dot M_*/ \dot M $ approaches unity does star formation inferfere with and possibly limit the amount of matter reaching the black hole.  There seem to be reasonable parameters that permit values of $\dot M$ as large as those powering the most luminous quasars.   

Although our estimated star formation rates depend on a number of highly uncertain parameters, certain trends are clear.  We find that little to no fragmentation should occur inside the black hole's gravitational radius of influence $r_{\rm BH}$ for $\dot M$ less than a few percent of a solar mass yr$^{-1}$, corresponding to $L \sim 10^{44}-10^{45}$ erg s$^{-1}$.  As $\dot M$ increases, star formation first sets in near or somewhat outside $r_{\rm BH}$, and migrates inward to radii $R_{\rm frag} \propto \dot M^{-0.76}$.  The total star formation rate within $r_{\rm BH}$ is dominated by conditions near the fragmentation radius, and increases steeply with the accretion rate $\propto \dot M^{1.71}$, until it becomes of the same order.  Thus, star formation in the equatorial zone of a magnetically elevated disk could produce a substantial annulus of stars inside the sphere of influence, the mass of which could be used to calibrate the luminosity and duration of a prior episode of activity.  

We can apply our model to the disk of young stars in the Galactic Center (GC), which has a rather well-defined inner edge at $\sim 1\arcsec$\citep{paumard06}, corresponding to $1.2 \times 10^{17}$ cm at a distance of 8 kpc. For a black hole mass $M = 4 \times 10^6 M_\odot$, this is well within the sphere of influence $r_{\rm BH} = 5.3 \times 10^{18} \sigma_{100}^{-2}$ cm, normalizing the velocity dispersion in the GC to 100 km s$^{-1}$.  If we identify the inner edge of the stellar disk with $R_{\rm frag}$, we obtain an accretion rate for the episode that formed the stars
\beq
\label{mdotGC}
\dot M_{\rm GC} \sim 0.2 \xi_{-1}^{32/29} \left({\alpha_{0.3} \over Q_{\rm crit} }\right)^{40/29} \ \msun \ {\rm yr}^{-1},
\eeq
where $\xi_{-1} = \xi / 0.1$.  For an accretion efficiency of 0.1 and all fiducial parameters equal to one, this would give a luminosity $\sim 10^{45}$ erg s$^{-1}$, corresponding to twice the Eddington limit for the GC black hole but, as we noted earlier, there are indications that $Q_{\rm crit}$ might be considerably larger in a strongly magnetized disk \citep{riols16}, which would depress $\dot M_{\rm GC}$. (We also stress that $\dot M_{\rm GC}$ estimated here is a mass supply rate far from the black hole; the question of how much mass actually reached the black hole is separate.)

Using this accretion rate, we estimate a star formation rate 
\beq
\label{MdotstarGC}
\dot M_{\rm *GC} \sim 6.5 \times 10^{-2} (\eta_{-2} E_{51} f_{-1})^{-1} \xi_{-1}^{0.42} \alpha_{0.3}^{0.52} Q_{\rm crit}^{-2.62}  \ \msun \ {\rm yr}^{-1}. 
\eeq 
For a Salpeter IMF, the total mass in stars formed during this accretion episode was $\sim 1.5 \times 10^4 M_\odot$, implying that the duration of the episode was 
\beq
\label{tGC}
t_{\rm *GC} \sim 2.3 \times 10^5 (\eta_{-2} E_{51} f_{-1}) \xi_{-1}^{-0.42} \alpha_{0.3}^{-0.52} Q_{\rm crit}^{2.62}  \ {\rm yr} , 
\eeq
which, on the surface, is too short for supernova feedback to come into equilibrium. However, this expression is very sensitive to the value of $Q_{\rm crit}$, and if we adopt $Q_{\rm crit} = 3$ (but all other fiducial parameters equal to one), then $t_{\rm GC}$ rises to 4 Myr, closer to the lifetime of a pre-supernova star.  The evolutionary state of the GC stars suggests that they were formed within $\sim 1 - 2$ Myr, implying that supernova feedback was not fully developed during this episode of accretion and allowing for a shorter burst of accretion with a higher star formation rate.  We note that if the IMF was very top-heavy, as suggested by \cite{nayakshin05}, then the increase in $t_{\rm *GC}$ implied by a larger value of $\eta_{-2}$ would be compensated by the smaller total mass in stars present in the disk.     
  
Magnetically elevated disk accretion is not expected to promote very large enhancements of star formation outside the sphere of influence. A growing body of observational evidence supports a correlation between star formation and AGN luminosity (or $L/L_\edd$), at rates that can reach $\sim 10^3$ times the accretion rate  \citep{bonfield11,mullaney12,chen13,delvecchio15, bernhard16}. ALMA data suggest that the surface star formation rate can be highly concentrated \citep{wilson14,oteo16}, although current resolution does not extend to within a few hundred parsecs. Extrapolation of our star formation estimates to $R > r_{\rm BH}$, however, suggest that $v_{\rm t}$ levels off, and the integrated star formation rate increases only logarithmically with radius.  A number of factors could render the magnetically elevated disk model --- and our quasi-local star formation scenario --- irrelevant at large $R$ (even if it applies at smaller radii), including insufficient poloidal magnetic flux, dominance of other mechanisms promoting accretion, such as global self-gravity or low intrinsic angular momentum of the gas supply, or feedback from the AGN itself \citep{silk13}. 

We stress that this feedback could be positive, enhancing star formation on large scales through several possible mechanisms.  For example, the bow shock associated with a jet could pressurize and compress a ring of gas where the jet blasts through, triggering star formation \citep{gaibler12}. A similar phenomenon holds for the more common nuclear AGN-driven winds in radio-quiet QSOs \citep{wagner13}. Widespread compression leading to induced star formation could also be driven by the interaction of the backflow from a jet cocoon with gas clumps in the inner kpc \citep{antonuccio10}. The latter phenomenon can self-regulate to give the canonical ratio $\sim 10^3$ of star formation rate to AGN accretion rate \citep{antonuccio16}, and may also apply for less collimated flows.  Indeed, observations confirm a connection between AGN-driven nuclear winds detected in X-ray absorption and fast molecular outflows (potentially hosting star formation) on large scales \citep{tombesi15}.  These mechanisms can operate concurrently with effects due to magnetic elevation in the inner disk, that serve to reduce $Q$ and regulate star formation inside $r_{\rm BH}$ to values considerably less than the AGN accretion rate. 

Unfortunately, the association of global star formation with AGN activity may be hard to disentangle from magnetically-regulated processes occurring closer in. Once a substantial burst of star formation is triggered, the rapid formation of massive stars and supernovae, and associated negative feedback on gas feeding, will mask any correlation between the AGN fueling process --- and associated nuclear star formation --- and high rates of star formation on larger scales \citep{pitchford16}, in part also because of the different duty cycles associated with these phenomena. The best indication of magnetically elevated accretion may well be relic stellar disks that persist after the activity has subsided, as in the GC.
  
We caution that this simple model for magnetically elevated disks represents an extrapolation from local (shearing box) simulations \citep{bai13,salvesen16a}.  The few global simulations that  demonstrate dynamo activity have generally agreed with the shearing box results, but involve either no imposed vertical field \citep{beckwith11,oneill11} or a weak vertical field \citep{suzuki14}, and therefore do not sample the parameter space of strongly magnetized disks.  Even the local simulations have not reached equatorial magnetizations beyond $\beta_0 \sim 0.3$, although we suggest that $\beta_0 \sim 0.1$ is attainable, and perhaps natural.  Given that elevated disks seem to require the presence of a strong vertical field \citep{salvesen16b}, it is important to understand how this field is either accumulated from the environment or generated locally through stochastic processes \citep[e.g.,][]{begelman14}.  The presence of poloidal flux presents the possibility that a significant fraction of the liberated energy goes directly into a magnetocentrifugal wind \citep{blandford82}, which we have not taken into account.  Evidence for fluctuating winds has been seen in various local simulations \citep[e.g.,][]{suzuki09,bai13,fromang13}, although these do not have the capability to model centrifugal effects, which are intrinsically global.  Moreover, if the disk height extends to values of $\xi = H/R \sim$ a few tenths, there is a real possibility that buoyancy of the toroidal field also contributes importantly to driving a wind \citep{contopoulos95}. 
 
The applications to AGN accretion discussed here depend on additional physical processes that are moderately well-understood in the hydrodynamical limit but which have hardly been studied in the highly magnetized (low-$\beta$) limit. Specifically, little research has been done on thermal instability or gravitational fragmentation in strongly magnetized disks, nor on supernova feedback in a strongly magnetized medium subject to strong cooling.  The simulations needed to tackle these problems will not be cheap; however, significant progress should be possible with currently available computing resources in the near future.  

\section*{Acknowledgements}
MCB acknowledges support from NASA Astrophysics Theory Program grants NNX14AB37G and NNX14AB42G
and NSF grant AST-1411879, and thanks the Institut d'Astrophysique de Paris and the Institut Lagrange de Paris for their hospitality and support.
JS was supported in part 
by ERC Project No. 267117 (DARK) hosted by Universit\'e Pierre et Marie Curie  (UPMC), Paris 6. JS also acknowledges the support of the JHU by NSF grant OIA-1124403.
We thank Phil Armitage for helpful discussions.

\bibliographystyle{mn2e}
\bibliography{../../biblio}

\end{document}